\documentstyle[12pt]{article}
\begin{document}
\centerline{\hfill IFIC/04-67}
\vskip0.5cm
\centerline{\Large\bf Magnetic helicity  and cosmological magnetic field}
\vskip0.5cm
\centerline{V.B. Semikoz $^{a,b}$ and D.D. Sokoloff $^{c}$}
\vskip0.5cm
\noindent
\mbox{
\hskip1cm$^{a}$ AHEP Group, Instituto de Fisica Corpuscular;
CSIC/Universitat}\\
\mbox{\hskip1cm{}de Valencia, Edificio de Paterna, Apartado 22085,
E-46071 Valencia, Spain,}\footnote{semikoz@ific.uv.es} \\
\mbox{\hskip1cm$^b$ IZMIRAN, Troitsk, Moscow Region 142190,
Russia,}\footnote{semikoz@orc.ru}\\
\mbox{\hskip1cm$^c$ Department of Physics, Moscow State University,
Moscow 119992, Russia}\footnote{sokoloff@dds.srcc.msu.su}
\vskip0.3cm
\centerline{\bf Abstract}
\vskip0.3cm
The magnetic helicity has paramount significance in
nonlinear saturation of galactic dynamo. We argue that the
magnetic helicity conservation is violated at the lepton stage in
the evolution of early Universe. As a result, a cosmological
magnetic field which can be a seed for the galactic dynamo obtains
from the beginning a substantial magnetic helicity which has to be
taken into account in the magnetic helicity balance at the later
stage of galactic dynamo.
\vskip0.3cm

\section{Introduction}

Magnetic fields of galaxies are believed to be generated by a
galactic dynamo based on the joint action of the so-called
$\alpha$-effect and differential rotation. The $\alpha$-effect is
connected with a violation of mirror symmetry in MHD-turbulence
and therefore caused by rotation. For a weak galactic magnetic
field, the mirror asymmetry is associated with helicity of the
velocity field and is proportional to the linkage of vortex lines.

The magnetic helicity is an inviscid integral of motion and its
conservation strongly constrains the nonlinear evolution of the
galactic magnetic field. The helicity density ${\bf B \cdot A}$ of
a galactic large-scale magnetic field is enhanced by galactic
dynamo (here ${\bf B}$ is a large-scale magnetic field, ${\bf A}$
is its vector-potential). Because of magnetic helicity
conservation, this income must be compensated by magnetic helicity
of a small-scale magnetic field. Note that the magnetic helicity
density is bounded from above by $b_l^2 l$, where $b_l$ is the
magnetic field strength at the scale $l$. Hence the supply of a
small-scale magnetic helicity occurs to be insufficient for the
compensation required. This fact strongly constrains the galactic
dynamo action (see \cite{Brandenburg,Brandenburg1} and references
therein).

On the other hand, a weak almost homogeneous cosmological magnetic
field $B_c$ can introduce a new element in this scheme. If we
adopt a thickness of the gaseous galactic disc $l_{\rm gal}\approx
1\,{\rm kpc}$ as a typical scale for galactic dynamo action  and
$B_c = 10^{-9}\,G$, then the magnetic helicity supplied via the
cosmological magnetic field ($h_c\simeq B_c^2l_H\sim 10^{10}~{\rm
G}^2\,{\rm cm}$, where $l_H$ is the horizon size) can be of the
same order as that one for the galactic magnetic field ($h_{\rm
gal}\simeq B_G^2l_{\rm gal}\sim 3\times 10^9\,{\rm G}^2{\rm cm}$,
$B_G\approx 10^{-6}\,{\rm G}$).

The cosmological magnetic field $B_c$ if it exists must be
substantially weaker than the galactic magnetic field (see \cite{Beck}
for review). According to the analysis of rotation
measures of remote radio sources, $B_c \le 10^{-10 ... -11}\,{\rm
G}$. This estimate however is based on the assumption that a
substantial part of charged particles in the Universe is in form
of thermal electrons in the intergalactic medium. A more robust
estimate, $B_c < 10^{-9} \, G$ is based on the isotropy
constrains.

We estimated above that if the cosmological magnetic field is of
the order $B_c = 10^{-9}\,G$ its magnetic helicity density can be
comparable with the magnetic helicity density of the galactic
magnetic field. Of course, the estimate $B_c = 10^{-11}$ gives a
much lower value for the magnetic helicity density. However it is
more than natural to expect a magnetic helicity concentration in
the processes of galactic formation. The question is whether a
mechanism of magnetic helicity production can be suggested for
physical processes in the early Universe.

Here we suggest a mechanism for magnetic helicity generation by a
collective neutrino-plasma interactions in the early Universe
after the electroweak phase transition.

\section{Weak force contribution to dynamo action in early Universe}

Let us consider the electron-positron plasma as a two-component
medium, for which the strong correlation between opposite charges
due to Coulomb forces gives ${\bf V}_\pm \approx {\bf V}$. Here
${\bf V}$ is the common fluid velocity of  {\it electroneutral}
conducting gas while its positively and negatively charged
components have different velocities, ${\bf V}_\pm ={\bf V}\pm
\delta {\bf V}$, where a small difference $\delta {\bf V}\ll {\bf
V}$ gives the separation of charges at small scales and enters the
electromagnetic current ${\bf j}^{(em)}= 2\mid e\mid n_{e}\gamma
_{e}\delta {\bf V}$ obeying in MHD the Maxwell equation, $4\pi
{\bf j}^{\rm (em)}=\nabla \times {\bf B}$ ($\hbar =c=1)$.

The electric field ${\bf E}$ derived from Euler equations for
plasma components takes the form \cite{Semikoz} which includes
the contribution of weak interactions ${\bf E} _{\rm weak}$ taken
in the collisionless Vlasov approximation. We do not consider
other known terms which describe weak interaction collisions
\cite{Dolgov} , Biermann battery effects, etc., and
which do not play substantially in favor of helicity generation.
In turn, we keep in ${\bf E}_{\rm weak}$ only the axial vector
term which violates the parity:

\[
{\bf E}_{\rm weak}^{(A)}=-\frac{G_{F}}{\sqrt{2}\mid \! e \! \mid
n_{e}}\sum_{\nu _{a}}c_{A}^{(a)}(n_{0}^{(-)}+n_{0}^{(+)})\hat{{\bf
b}}~\frac{\partial \delta n_{\nu _{a}}({\bf x},t)}{\partial t}+
\]

\begin{equation}
+(N_{0}^{(-)}+N_{0}^{(+)})\nabla (\hat{{\bf b}}\cdot \delta {\bf j}%
_{\nu _{a}}({\bf x},t))~.  \label{axialelectric}
\end{equation}%
Here $G_{F}=10^{-5}/m_p^2$ is the Fermi constant, $m_{p}$ is the
proton mass; $c_{A}^{(a)}=\mp 0.5$ is the axial weak coupling,
upper (lower) sign is for electron (muon or tau) neutrinos,
$\delta n_{\nu _{a}}=n_{\nu _{a}}-n_{\overline{\nu }_{a}}$is the
neutrino density asymmetry, $\delta {\bf j}_{\nu _{a}}={\bf
j}_{\nu _{a}}-{\bf j}_{\overline{\nu }_{a}}$ is the neutrino
current asymmetry; $\hat{{\bf b}}$ is the unit vector along the
mean magnetic field; $n_{0}$ is the lepton number density at the
main Landau
level given by the equilibrium Fermi distribution in a hot plasma, $%
n_{0}^{(-)}\approx n_{0}^{(+)}=(\mid e\mid B/2\pi ^{2})T\ln 2,$ where at
equilibrium the temperature obeys $T_{+}=T_{-}=T\gg m_{e}$ $.$ In the
non-relativistic limit, $\rm{v}\ll 1$, the relativistic polarization
terms \ in Eq. (\ref{axialelectric}) tend to the lepton densities at the
main Landau level, $N_{0}^{(\sigma )}\rightarrow n_{0}^{(\sigma )}$ $.$

Accounting for the first line in Eq. (\ref{axialelectric}) we
obtain the axial vector term ${\bf E}_{\rm weak}^{(A)}=-\alpha
{\bf B,}$ where the helicity coefficient $\alpha $ is the  {\it
scalar }in the standard model (SM) with neutrinos instead of the
pseudoscalar $\langle \bf{v}\cdot (\nabla \times {\bf v})\rangle
)$ in standard MHD:

\[
\alpha =\frac{G_{F}}{2\sqrt{2}\mid e\mid B}\sum_{a}c_{A}^{(\nu _{a})}\left[
\left( \frac{n_{0}^{(-)}+n_{0}^{(+)}}{n_{e}}\right) \frac{\partial \delta
n_{\nu _{a}}}{\partial t}\right] \simeq
\]

\begin{equation}
\simeq \frac{\ln 2}{4\sqrt{2}\pi ^{2}}\left( \frac{10^{-5}T}{%
m_{p}^{2}\lambda _{\rm{fluid}}^{(\nu )}}\right) \left( \frac{\delta
n_{\nu }}{n_{\nu }}\right) ~.  \label{alpha}
\end{equation}

Here we substituted $n_{\nu }/n_{e}=0.5$, and assumed a scale of
neutrino fluid inhomogeneity $t\sim \lambda _{\rm{fluid}}^{(\nu
)}$, that is small compared with a large $\Lambda $-scale of the
mean magnetic field.

Let us stress that instead of the  {\it  difference} of electron
and positron contributions in axial vector terms entering the pair
motion equation \cite{Semikoz} and given by the polarized density
asymmetries $\sim (n_{0}^{(-)}-n_{0}^{(+)})$ we obtained here the  {\it  sum%
} of them $\sim (n_{0}^{(-)}+n_{0}^{(+)})$ that can lead to a
significant effect in a hot plasma.

The admixture of the pseudovector $\alpha {\bf B}$ to the pure
vector ${\bf E}$, e.g.  for the constitutive relations ${\bf D}=
\varepsilon {\bf E} + \beta {\bf B}$, ${\bf H}= \gamma {\bf E} +
\mu^{-1}{\bf B}$ due to the same neutrino-plasma weak interaction
described by the constants $\beta$, $\gamma$, has been already
discussed in literature (see \cite{Nieves}, Eqs. (3.5),
(3.6)).  In a forthcoming paper we show that such unusual
coefficients appear in {\it chiral media} and are simply connected
with $\alpha$ given by Eq.  (\ref{alpha}), $(\varepsilon
-1)\alpha= \beta + \gamma$, where $\varepsilon$ is the dielectric
permittivity of plasma.

Thus, using Eq. (\ref{axialelectric}) \ from the Maxwell equation $\partial _{t}%
{\bf B}=-\nabla \times {\bf E}$ one obtains the Faraday equation
generalized in SM with neutrinos and antineutrinos:

\begin{equation}
\frac{\partial {\bf B}}{\partial t}=\nabla \times \alpha
{\bf B}+\eta \nabla ^{2}{\bf B}~,  \label{Faradey2}
\end{equation}
where we omitted the weak vector contribution $\nabla \times {\bf
E} _{\rm weak}^{(V)}\sim \nabla \times \delta {\bf j}_{\nu
_{a}}({\bf x},t)$ suggested by Brizard et al (2000) since we
neglect any neutrino flux vorticity in the hot plasma of early
universe. Because the early universe is almost perfectly
isotropisc and homogeneous, we ignore here any contribution from
large-scale motions as well. In the relativistic plasma the
diffusion coefficient $\eta $ takes the form $\eta =(4\pi \times
137$ $T)^{-1}.$

The first term in r.h.s. of Eq. (\ref{Faradey2}), $\nabla \times \alpha
{\bf B}$, is associated with the parity violation in weak interactions
in the early universe plasma.

We stress that the Eq.~(\ref{Faradey2}) is the usual equation for
mean magnetic field evolution with $\alpha $-effect based on
particle effects rather on the averaging of turbulent pulsations.
It is well-known (see e.g. \cite{Ruzmaikin}) that Eq.~(\ref
{Faradey2}) describes a self-excitation of a magnetic field with
the spatial scale $\Lambda \approx \eta /\alpha $ and the growth
rate $\alpha ^{2}/4\eta $. Authors \cite{SS}
estimated these values for the early universe to get

\begin{equation}  \label{scale}
\frac{\Lambda}{l_H} = 1.6\times
10^9\left(\frac{T}{\rm{MeV}}\right)^{-5}
\left(\frac{\lambda^{(\nu)}_{\rm fluid}}{l_{\nu}(T)}\right) (\mid
\xi_{\nu_e}(T)\mid)^{-1}~,
\end{equation}

\begin{equation}
B(x)=B_{\max }\exp \left( 25\int_{x}^{1}\left( \frac{\xi _{\nu _{e}}({\it
x}^{\prime })}{0.07}\right) ^{2}{\it x'}^{10}d{\it x}^{\prime }%
\right) \,.  \label{last}
\end{equation}%
Here $l_{H}$$(T)=(2 \cal {{H})}$$^{-1}$ is the horizon size and $
\cal {H}$=$4.46\times 10^{-22}(T/\rm{MeV})^2~\rm {MeV}$ is the
Hubble parameter; the variable ${\it x}=T/2\cdot 10^{4}\rm{MeV}$
corresponds to the maximum temperature $T\simeq 20~\rm{GeV}\ll
T_{EW}\simeq 100~\rm{GeV}$ for which the point-like Fermi
approximation for weak interactions we rely on is still valid.
Finally $l_{\nu}(T)$ is the neutrino free path and
$\xi_{\nu_e}=\mu_{\nu_e}/T\ll 1$ is the small dimensionless
electron neutrino chemical potential normalized in (\ref{last}) on
the maximum value $\xi_{\nu_e}\leq 0.07$ allowed by the Big Bang
Nucleosynthesis (BBN) bound on light element abundance \cite{Dolgov1}.

Thus, while in the temperature region $T_{EW}\gg T\gg T_{0}=10^{2}~\rm{%
MeV}$ there are many small random magnetic field domains, a weak
mean magnetic field turns out to be developed into the uniform
{\it  global} magnetic field at temperatures below $T_0$ (see Eq.
(\ref{scale}).  The global magnetic field can be small enough to
preserve the observed isotropy of the cosmological model
\cite{Zeld} while being strong enough to be interesting as a
seed for galactic magnetic fields.  This scenario was extensively
discussed by experts in galactic magnetism \cite{Kulsrud},
however until now no viable origin for the global magnetic field
has been suggested. We believe that the dynamo based on the
$\alpha $-effect induced by particle physics solves this
fundamental problem and opens a new and important option in
galactic magnetism.

\section{Magnetic helicity generation by collective neutrino-plasma
interactions}

Let us consider how the collisionless neutrino interaction with
charged leptons can produce the  primordial magnetic helicity
$H=\int_v({\bf A}\cdot {\bf B})d^3x$, where $v$ is the volume that
encloses the magnetic field lines.

For that we should substitute into the derivative,

\begin{equation}
\frac{d\,H}{d\,t}=-2\int_{v}({\bf E}\cdot {\bf B}%
)d^{3}x~,  \label{helicity1}
\end{equation}
the electric field ${\bf E}$ given by Eq. (\ref{axialelectric}).
Neglecting any rotation of primordial plasma given by the first
dynamo term, or retaining  the resistive term and the weak
interaction term ${\bf E}_{\rm weak}^{(A)}$ given by Eq.
(\ref{axialelectric}) that is the main one in the absence of any
vorticities, one finds from (\ref {helicity1})

\begin{equation}
\frac{d\,H}{d\,t}=-2\eta \int_{v}d^{3}x(\nabla \times {\bf
B})\cdot {\bf B}+2\alpha \int_{v}d^{3}xB^{2}.
\label{helicitychange}
\end{equation}

Note that the second term in the r.h.s. violates parity: it is a
pure {\it  scalar }while other terms are pseudoscalars as it
should be for the helicity $\rm H$ in standard MHD. Nevertheless,
all terms in the generalized helicity evolution equation
(\ref{helicitychange}) obey $CP$-invariance as it should be for
the electroweak interactions in SM since the new coefficient
$\alpha $ (\ref{alpha}) is $CP$-odd, $(CP)\,\alpha~(CP)^{-1}
=-\alpha $, as well as $\nabla \times ...$. This is due to the
changes $n_{0-}\longleftrightarrow n_{0+}$ and $\delta n_{\nu
_{a}}\rightarrow -\delta n_{\nu _{a}}$ in (\ref{alpha}), provided
by the well-known properties: particle helicities are $P$-odd and
particles become antiparticles under the charge conjugation
operation $C$, in particular, active left-handed neutrinos convert
to the active right-handed neutrinos under $CP$-operation, $\nu
_{a}\longrightarrow \overline{\nu }_{a}.$ Obviously, the product
(${\bf E}\cdot {\bf B)}$ entering the helicity evolution
(\ref{helicity1}) is $CP$-odd too because both electric and
magnetic fields are $C$-odd and have opposite $P$-parities.

First term in the r.h.s. of Eq.~(\ref{helicitychange}) gives
conventional ohmic losses for magnetic helicity and usually is
neglected in helicity balance.

\section{Seed magnetic helicity in cosmology}
Neglecting in evolution equation (\ref{helicitychange}) the first
diffusion term, we can calculate the magnetic helicity $H(t)$
using Eq.~(\ref{last}) to yield

\begin{equation}\label{cosmhel}H(t)=2B_{\rm{max}}^2\int\limits_v
\!\!\! d^3r\!\!\!\!\int\limits_{t_{\rm{max}}}^t
\!\!\!\!dt'\alpha(t')\, e^{
\int\limits_{t_{\rm{max}}}^{t'}\left[\frac{\alpha^2(t^{''})}{4\eta
(t^{''})}\right]dt^{''}}\!\!\!\!\!\! + H(t_{\rm{max}}),
\end{equation}
where $H(t_{\rm{max}})$ is the initial helicity value at the
moment $t_{\rm{max}}$ if it exists, and we present the helicity
density entering the integrand as $h(x)=2.4\times
10^3(B_{\rm{max}})^2m_e^{-1}J(x)$. Here we use the dimensionless
variable $x=T/2\cdot 10^4~\rm{MeV}$. The maximum value $x=1$
corresponds to the maximum temperature $T_{\rm{max}}\simeq
20~\rm{GeV}\ll T_{EW}\sim 100~\rm{GeV}$ (see motivation above),
and we assumed that the neutrino gas inhomogeneity scale is of the
order of the neutrino free path $l_{\nu}(T)$,
$\lambda_{\rm{fluid}}^{(\nu)}\sim l_{\nu}$. The WKB value of the
mean magnetic field amplitude $B_{\rm{max}}$ obeys
$B_{\rm{max}}\ll T_{\rm{max}}^2/e=4(T_{\rm{max}}/\rm{MeV})^2B_c$,
where the Schwinger field $B_c=m_e^2/e=4.41\times 10^{13}\,{\rm
G}$. Hence we may introduce a small WKB parameter $\kappa\ll 1$
for the mean field $B_{\rm{max}}=\kappa T_{\rm{max}}^2/e\ll
T^2_{EW}/e$, or substituting $T_{\rm max}=20~\rm{GeV}$ one obtains
$B_{\rm max}=\kappa\times 7\times 10^{22}\,{\rm G}\ll B_{EW}\sim
10^{24}\,{\rm G}$, where, let us say, $\kappa\sim 0.01$.

It is worth noting that such WKB value of $B_{\rm{max}}$ which is
scaled being frozen-in as $B(T)= B_{\rm{max}}(T/T_{\rm{max}})^2$
does obey the BBN limit $B \le 10^{11}~{\rm G}$ at the temperature
$T_{\rm{BBN}}\simeq 0.1~\rm{MeV}$, i.e.
$B(T_{\rm{BBN}})=\kappa\times (7/4)\times 10^{12}~{\rm{G}}$. Note
also that the sign of the first term in (\ref{cosmhel}) is not
well determined since it depends on the combined neutrino density
asymmetry \cite{SS}, $\delta n_{\nu}/n_{\nu}=
\sum_a c^{(A)}_{e\nu_a}\delta n_{\nu_a}/n_{\nu_a}\sim
[\xi_{\nu_{\mu}} + \xi_{\nu_{\tau}} - \xi_{\nu_e}]$, where the
values of the dimensionless neutrino chemical potentials
$\xi_{\nu_a}=\mu_{\nu_a}/T$ are given by the BBN limit cite{Dolgov1}
: $-0.01< \xi_{\nu_e}< 0.07$, or by the CMBR/LSS bound
\cite{Hansen}: $-0.01< \xi_{\nu_e}< 0.22$, $\mid
\xi_{\nu_{\mu, \tau}}\mid < 2.6$. One can use, e.g., the
conservation of the lepton number $L_e -L_{\mu}$ that implies
$\xi_{\nu_e}=- \xi_{\nu_{\mu}}$, however, this does not guarantee
the definite sign of the combined neutrino density asymmetry
$\delta n_{\nu}/n_{\nu}$. The definite sign of the magnetic
helicity (left-handed, $H<0$), arising during electroweak
baryogenesis \cite{Vachaspati} is another case connected with the
CP-violation.

Let us emphasize that cosmological helicity production via the
collective neutrino interaction with hot plasma ceases if the
neutrino chemical potential $\mu_{\nu_e}$ (hence the neutrino
density asymmetry $\delta n_{\nu}/n_{\nu}$) vanishes,
$\xi_{\nu_e}=\mu_{\nu_e}/T\to 0$, $\delta n_{\nu}/n_{\nu}\to 0$.
Solely the inequality $\xi_{\nu_e}\leq 0.07$ is known from the BBN
bound on light elements abundance at $T< O(\rm{MeV})$ (Dolgov et
al, 2002), thereby substituting for a rough estimate $\xi_{\nu_e}=
0.07$ we estimate the integral for $h(x)$ as $J(x)\sim 10$.

Thus, collecting numbers for $B_{\rm{max}}$, $J(x)$ and using the
electron Compton length $m_e^{-1}=3.86\times 10^{-11}~\rm{cm}$,
one finds the huge value of {\it cosmological helicity density }
that could seed galactic magnetic helicity

\begin{equation} \label{estimate} h(x)\simeq
4.5\times 10^{38}\kappa^2J(x)\,\rm{G}^2\rm{cm}\sim  4.5\times
10^{39}\kappa^2\,\rm{G}^2\rm{cm}.
\end{equation}

\section{Discussion}

Traditional galactic dynamo considered galactic magnetic field
produced from a very weak seed field. This implies that the
magnetic helicity of the seed field is weak. We argue that the
applicability of this viewpoint is limited. The seed field for
galactic dynamo can be a field of substantial strength and
substantial helicity. The first part of this statement is already
quite well-accepted in modern galactic dynamo (see e.g. \cite{Beck})
while the second one is new. In this letter we suggest a
physical mechanism for the magnetic helicity production for the
seed field of galactic dynamo. As far as we know, such mechanisms
was not considered previously.  Note that \cite{Field}
pointed out in a general form the importance of the electroweak
phase transition for magnetic helicity generation.

We stress that the epoch just after the electro-weak phase
transition and that one of galaxy formation are quite remote in
respect to their physical properties. We appreciate that the
magnetic helicity evolution in the time interval between these
epoches has to be addressed separately.  In particular,
large-scale magnetic helicity produced by galactic dynamo is
antisymmetric in respect to the galactic equator while the
magnetic helicity  from any cosmological sources is obviously
independent on the galactic equator position. It is far from clear
how important this asymmetry is for nonlinear galactic dynamos and
for the observed asymmetry of magnetic field in Milky Way. Note
also that the strong cosmological magnetic field could prevent the
inverse MHD cascade on the scale of galaxies \cite{Milano},
\cite{axel}.

We should remark that the huge helicity value (\ref{estimate})
exists only in hot ultrarelativistic ($T\gg m_e$) early universe
plasma where $\alpha (T)$ is sufficiently large. The evolution of
magnetic helicity ${\rm H}(t)$, or how cosmological magnetic
helicity feeds protogalactic fields is a complicated task. In the
nonrelativistic plasma, first, positrons vanish, then with the
cooling  for the frozen-in magnetic field $B\sim T^2\to 0$ the
electron density at the main Landau level drops, $n_{0-}\to 0$,
resulting in $\alpha\to 0$, and the magnetic helicity production
becomes impossible.

Let us note that the neutrino collision mechanism \cite{Dolgov} can not
produce magnetic helicity unlike in our collisionless  mechanism. This
immediately comes after the substitution of the electric field term
stipulated by weak collisions ${\bf E}=-\mid J_{\rm ext}\mid{\bf
V}/\sigma$ ($\sigma$ is the electric conductivity in the ultrarelativistic
plasma) and taken from Eq. (5) in \cite{Dolgov}, where the
electric current $J_{\rm ext}\sim G_F^2$ is caused by the friction force
due to the difference of weak cross-sections for neutrino
scattering off electrons and positrons. This current is directed
along the fluid velocity. The generalized momentum ${\bf
P}=w_e\gamma_e{\bf V}$, $w_e=4T$ is the enthalpy, $\gamma_e\gg 1$
is the $\gamma$-factor in the ultrarelativistic plasma, obeys
Euler equation \cite{Semikoz}

\begin{equation}
\left(\partial_t + {\bf V}\cdot \nabla\right){\bf P}=-\frac{\nabla
p}{n_e} + \frac{\rm{rot}~{\bf B}\times {\bf B}}{4\pi n_e} + {\rm
~weak~~ terms}\,, \label {sem04}
\end{equation}
from which retaining the standard MHD terms only (the first and
the second ones in the r.h.s. of Euler equation) one can obtain
the velocity ${\bf V}\propto \int_{t_0}^t[-\nabla p/n_e + ({\rm
rot}~{\bf B}\times {\bf B})/4\pi n_e]$ that does not contribute
(in the lowest approximation over $\sim G_F^2$) to the helicity
change.

Let us note that we rely here on homogeneous magnetic fields with
a scale which is less (however comparable) than the horizon $l_H$,
hence the magnetic force lines are closed within the integration
volume $\int_vd^3r(...)$, or applying the Gauss theorem one can show that
the contribution of the first term in the r.h.s.  of Eq.~(\ref{sem04})
to the helicity production (\ref{helicity1}) vanishes.  This is exactly
like for the gauge transformation of the vector potential ${\bf A}$ in the
helicity ${\rm H}=\int_vd^3r{\bf A}\cdot{\bf B}$, ${\bf A}\to {\bf A}+
\nabla \chi$. On the other hand, there remains an open question how to
define the gauge invariant helicity for superhorizon scales.

There are other astrophysical objects for which axial vector weak
forces acting on electric charges and driven by neutrinos can lead
to the amplification of mean magnetic field and its helicity as
given in (\ref{helicitychange}).  For instance, the neutrino flux
vorticity which is proportional to $\nabla \times {\bf j}_{(\nu
)}$ can vanish for isotropic neutrino emission from supernovas in
the diffusion approximation when neutrino flux  ${\bf j}_{(\nu
)}(r)$ is parallel to the radius $\bf r$.  In such case the
mechanism of collective neutrino-plasma interactions originated by
the axial vector weak currents becomes more efficient to amplify
magnetic field than the analogous mechanism based on weak vector
currents \cite{Brizard}.

\vskip0.3cm

{\bf Acknowledgements} This work was supported by the RFFI grants
04-02-16094 and 04-02-16386. Helpful discussions with
A.Brandenburg and T.Vachaspati are acknowledged.

\end{document}